\documentclass[%
reprint,
  amsmath,amssymb,
  aps,
prl,
]{revtex4-1}

\usepackage{graphicx}
\usepackage{dcolumn}
\usepackage{bm}
\usepackage{color}
\usepackage{ulem}
\usepackage{natbib}

\usepackage{blindtext}
\usepackage{subfiles}

\begin{document}

\title{Matrix viscoelasticity decouples bubble growth and dynamics in coarsening foams.}

\author{Chiara Guidolin$^{1,2*}$, Emmanuelle Rio$^1$, Roberto Cerbino$^3$, Fabio Giavazzi$^{2*}$, and Anniina Salonen$^{1*}$}
\affiliation{$^1$Université Paris-Saclay, CNRS, Laboratoire de Physique des Solides, Orsay, France.\\
$^2$Department of Medical Biotechnology and Translational Medicine, University of Milan, Segrate, Italy.\\
$^3$Faculty of Physics, University of Vienna, Vienna, Austria.\\
$^*$chiara.guidolin@unimi.it,
fabio.giavazzi@unimi.it,
anniina.salonen@universite-paris-saclay.fr}

\begin{abstract}
Pressure-driven coarsening triggers bubble rearrangements in liquid foams.
Our experiments show that changing the continuous phase rheology can alter these internal bubble dynamics without influencing the coarsening kinetics. Through bubble tracking, we find that increasing the matrix yield stress permits bubble growth without stress relaxation via neighbor-switching events, promoting more spatially homogeneous rearrangements and decoupling bubble growth from dynamics. This eventually leads to a structural change which directly impacts the foam mechanical and stability properties, essential for applications in various technological and industrial contexts.
\end{abstract}

\maketitle

\noindent

Liquid foams are soft materials made of gas bubbles tightly packed together in a continuous liquid medium.
Such systems can age through pressure-driven gas transfer between the bubbles, which results in a gradual growth of the mean bubble size \cite{VonNeumann1952}.
This coarsening process induces internal stress-driven dynamics: bubble size variations give rise to imbalanced stresses inside the foam, which eventually relax through local bubble rearrangements entailing an exchange of neighbours \cite{Cantat2013book}.
Probing the link between restructuring events and the coarsening process in model systems like foams can shed a light on the interplay between stress accumulation, local yielding, and structural relaxation which is common to a variety of other soft glassy systems \cite{Sollich1997}.

\begin{figure}[htbp]
    \centering \includegraphics[width=\columnwidth]{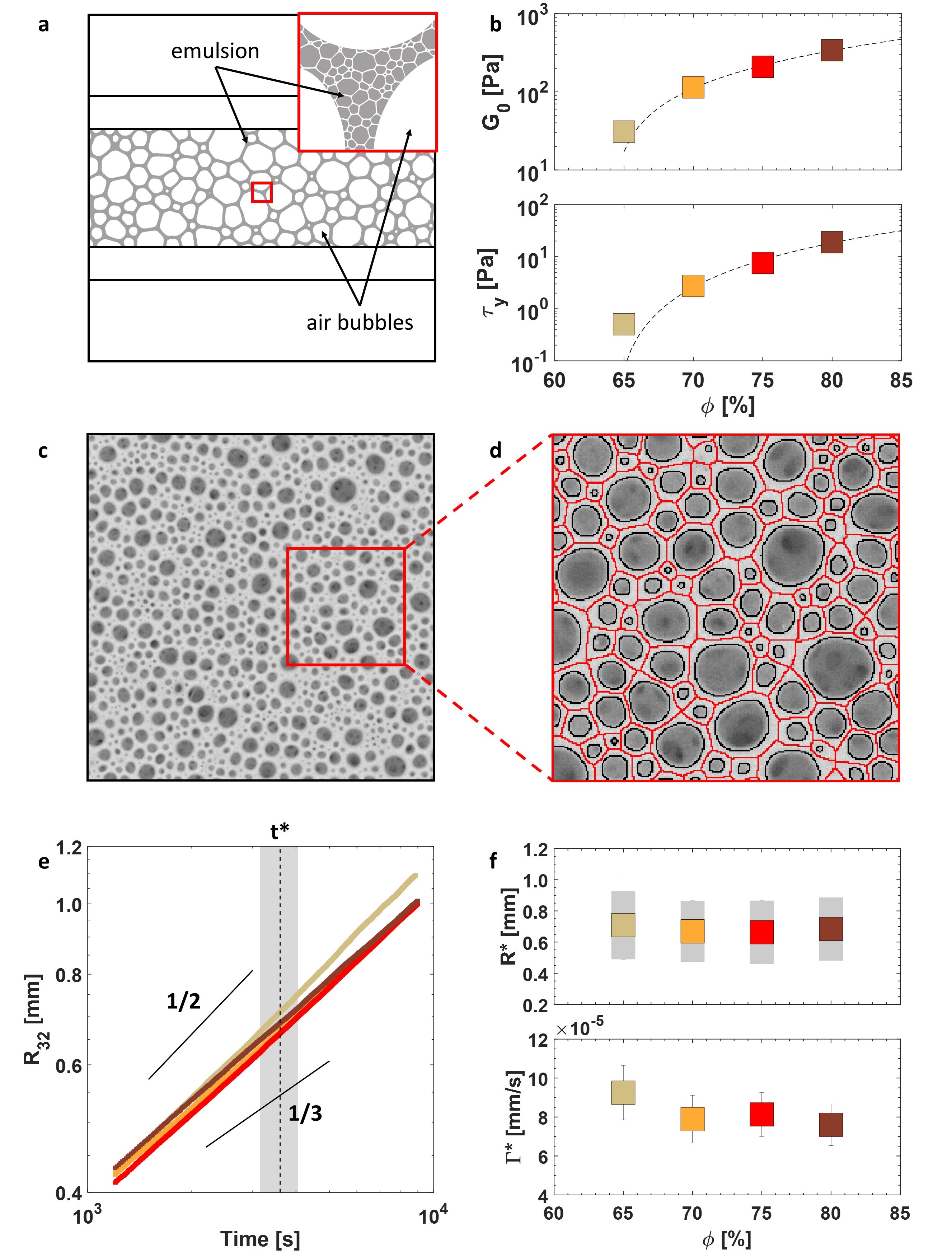}
    \caption{\textbf{Foam coarsening in concentrated emulsions.}
    (a) Illustration of the foam cell, highlighting the oil-in-water emulsion between the bubbles.
    (b) Emulsion storage modulus $G_0$ (top) and yield stress $\tau_y$ (bottom) at different $\phi$. Experimental details can be found in \cite{Guidolin2023}. The dashed lines represent the two expected scalings $G_0\sim\phi(\phi-\phi^*)$ \cite{Mason1995} and $\tau_y\sim(\phi-\phi^*)^2$ \cite{Mason1996}, where $\phi^*=64\%$.
    (c) Top view of a foam at $\phi$=75\% at $t^*=60$ minutes.
    The frame edge is 30 mm.
    (d) Enlargement showing the outline of the bubble segmentation (black) and the associated foam skeleton (red). The frame edge is 10 mm.
    (e) Mean bubble growth at different $\phi$.
    The gray-shaded area highlights the time window considered, centered around $t^*$= 60 minutes (vertical dashed line).
    (f) Mean bubble size and coarsening rate evaluated at $t=t^*$. The gray-shaded bars on $R^*$ represent the standard deviation of the bubble size distribution.}
    \label{fig:FIGURE_1}
\end{figure}

Bubble dynamics has been widely probed in coarsening aqueous foams with light scattering techniques \cite{Durian1991_Gillette, Gittings2008,LeMerrer2012PRL,Sessoms2010}.
The frequency of rearrangements is set by the strain rate imposed by the coarsening process, thus it is directly linked to the coarsening rate \cite{Sessoms2010}.
A combination of reciprocal and direct space analysis has more recently shown how coarsening dynamics is governed by directionally-persistent bubble displacements up to a critical length scale given by the bubble size \cite{Giavazzi2021}.
The loss of persistency has been ascribed to sudden changes in the local stress configuration due to the occurrence of plastic bubble rearrangements
inside the foam.

While all this holds for aqueous foams, many practical applications involve foams made from more complex fluids, like gels, emulsions, or pastes.
The behavior and characteristics of these types of foams are considerably less understood.
How easily bubbles can move during coarsening clearly depends on the amount, as well as on the nature of the medium between them \cite{LeMerrer2012PRL}.
Both surface and bulk rheology of the continuous phase significantly affect the duration of plastic events within the foam.
For instance, increased interfacial and bulk viscosity both translate into slower rearrangements \cite{LeMerrer2013PRE}.
By contrast, the effect of elastic properties has been scarcely investigated.
Conditions required for preventing or arresting coarsening have been proposed with high foam elasticity \cite{Bey2017} or continuous phase yield stress \cite{Lesov2014, Feneuil2019}.
However, a thorough description of the impact of bulk elasticity on coarsening dynamics is still missing.
Recently, spatially heterogeneous coarsening has been observed in foams with viscoelastic continuous phases, eventually leading to atypical foam morphologies \cite{Guidolin2023}.
Establishing the link between coarsening dynamics and structural evolution in complex foams would thus allow for a finer control of internal foam structure, which is crucial for applications, but still calls for a systematic investigation of how bubble motion changes with matrix stiffness.

In this Letter, we experimentally investigate how the presence of a viscoelastic medium between the bubbles affects their motility during coarsening.
By performing bubble tracking, we show that a gradual increase of the continuous phase yield stress dramatically restricts bubble motion during coarsening.
Moreover, we show that the marked change in the bubble dynamics is not mirrored by a significant change in the coarsening kinetics, suggesting a decoupling between the coarsening rate and the rate of restructuring events inside the foam.

We follow the ageing of foams made of concentrated oil-in-water emulsions, as sketched in Fig. \ref{fig:FIGURE_1}(a).
At oil volume fractions $\phi$ above random close packing, emulsions have a storage modulus and a yield stress both increasing with $\phi$ \cite{Mason1995,Mason1996}.
We consider oil fractions ranging between 65\% and 80\%, so that we vary the elastic modulus of the foam continuous phase $G_0$ between 30 and 340 Pa, and its yield stress $\tau_y$ from 0.5 to 20 Pa, as shown in Fig. \ref{fig:FIGURE_1}(b).
In this range of $\phi$, the emulsion yield stress delays the gravitational drainage \cite{Goyon2010_PRL_drainage} allowing the foam to coarsen at homogeneous liquid fraction (Supplemental Material, section ).

Emulsions are first prepared at the desired $\phi$ by mixing rapeseed oil (from Brassica rapa, Sigma Aldrich) and an aqueous surfactant solution (Sodium Dodecyl Sulphate 30 g/L, Sigma Aldrich) with the double-syringe method \cite{Gaillard2017}, and then foamed with the aid of a planetary kitchen mixer until the sample volume has increased tenfold (liquid fraction ranging between 9\% and 11\%).
The resulting foam is then gently sandwiched between two square glass plates (edge 20 cm), with a spacing (10 mm) much larger than the typical bubble size ($R \sim 10^{-1}$ mm) so that the foam sample can be safely considered three-dimensional.
The typical bubble Laplace pressure ($\gamma/R \simeq 300$ Pa with $\gamma \simeq 30$ mN/m) is higher than the interstitial emulsion yield stress, ensuring that our foams coarsen \cite{Lesov2014, Feneuil2019}.

\begin{figure*}[htb]
    \centering
    \includegraphics[width=\textwidth]{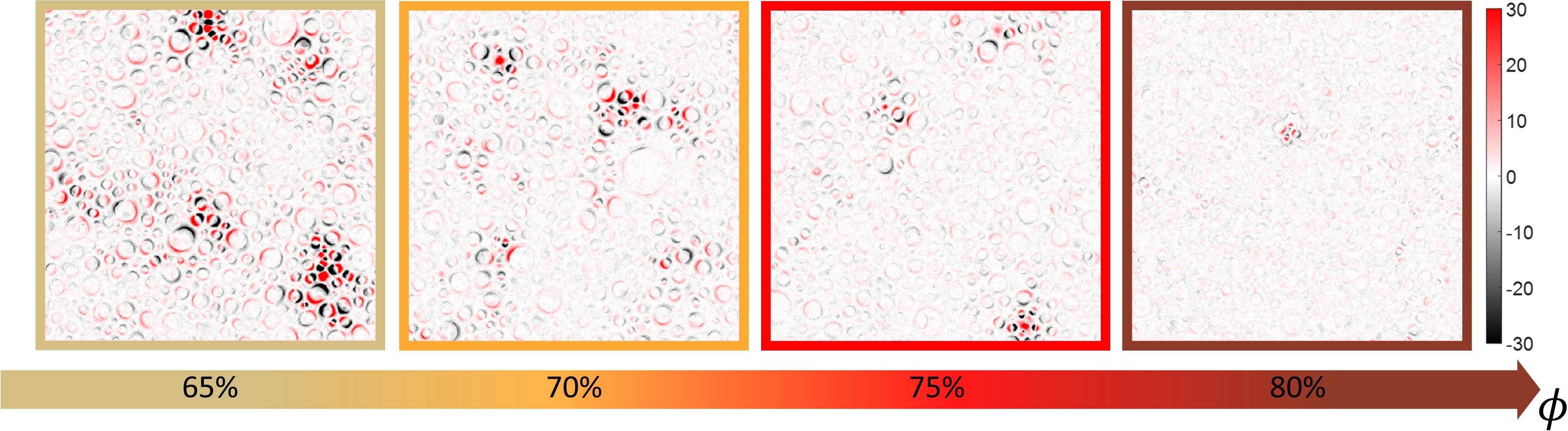}
    \caption{\textbf{Activity maps.} Representative activity maps of coarsening foams at increasing $\phi$.
    The maps are obtained as the difference between two frames separated by $\Delta t$ = 60 s. The sample activity is compared at the same foam age $t=t^*$ for each sample. The frame edge is 25 mm.}
    \label{fig:FIGURE_2}
\end{figure*}

Foam ageing is monitored from the top by taking image stacks with a camera (Basler acA3800-14um, equipped with a Tamron lens 16mm F/1.4), while a square array of LED lights provides uniform illumination from above.
The typical foam appearance is shown in Fig. \ref{fig:FIGURE_1}(c).
Image processing is performed with custom MATLAB scripts as follows.
Foam pictures are first segmented through an adaptive thresholding and then skeletonized with a watershed algorithm.
An example of bubble segmentation and foam skeleton outline is shown in Fig. \ref{fig:FIGURE_1}(d).
From the foam skeleton, we estimate the size of each bubble as the radius $R=\sqrt{A/\pi}$, where $A$ is the area of the polygonal cell, to then monitor how its average value, taken as $R_{32}=\langle R^3\rangle / \langle R^2\rangle$, evolves over time.
The time evolution of $R_{32}$ at different $\phi$ is reported in Fig. \ref{fig:FIGURE_1}(e): despite the tenfold increase in the emulsion elasticity, the mean bubble growth is approximately the same for each sample over almost one decade in time.
At the liquid fractions considered ($\sim$ 10\%), neighbouring bubbles share thin liquid films, whose surface area is however reduced by the presence of thick Plateau borders.
Inter-bubble gas diffusion is thus slower than in a dry foam (liquid fraction $\sim 1\%$) with the same skeleton \cite{Schimming}.
The experimental curves of $R_{32}(t)$ indeed lie between the two power laws predicted for the mean bubble growth in very dry foams \cite{Mullins1986} and dilute bubbly liquids \cite{Lifshitz1961} in their scaling state (SM, Fig. S2).
However, in contrast to aqueous foams, the foams under study are not expected to head towards self-similarity.
In fact, lack of liquid phase redistribution and a non-trivial interplay between foam structure and rheology has been shown to eventually lead to highly heterogeneous structures at millimetric bubble sizes \cite{Guidolin2023}.

We now compare the coarsening dynamics at the same foam age $t^*$.
We present here data for $t^*=$ 60 minutes, but choosing different values of $t^*$ does not affect the final results (SM, section 1). The mean bubble size $R^*=R_{32}(t^*)$ and its growth rate $\Gamma^*=$ d$R_{32}/$d$t|_{t^*}$ show no significant dependence on $\phi$, as shown in Fig. \ref{fig:FIGURE_1}(f).
The bubble size distribution also does not change with $\phi$ (SM, Fig. S3).

However, a visual inspection of the coarsening movies reveals a dramatic change in the global bubble dynamics with increasing $\phi$ (SM, movies SM1-SM4).
At $\phi$=65\%, bubbles freely rearrange as they coarsen, as in an aqueous foam.
By contrast, as $\phi$ is increased up to 80\%, their motility is drastically reduced: bubbles appear to just grow or shrink while remaining substantially stuck in their initial positions.

The different bubble dynamics can be better appreciated with the aid of activity maps, like the ones shown in Fig. \ref{fig:FIGURE_2}.
The higher activity of the foam at $\phi$=65\% stands out at first glance.
Here, we recognize the occurrence of topological changes as the regions where two couples of red and black sickle areas emerge symmetrically, marking the neighbor-switching event.
Furthermore, these plastic events cause substantial motion in the surrounding bubbles, extending up to 4-5 bubbles away, as in aqueous foams \cite{Gittings2008, Sessoms2010}.
On the other hand, as $\phi$ is increased, bubble rearrangements rarefy (SM, Fig. S8-S9) and the overall activity is drastically reduced.

\begin{figure}[htb]
    \centering
    \includegraphics[width=\columnwidth]{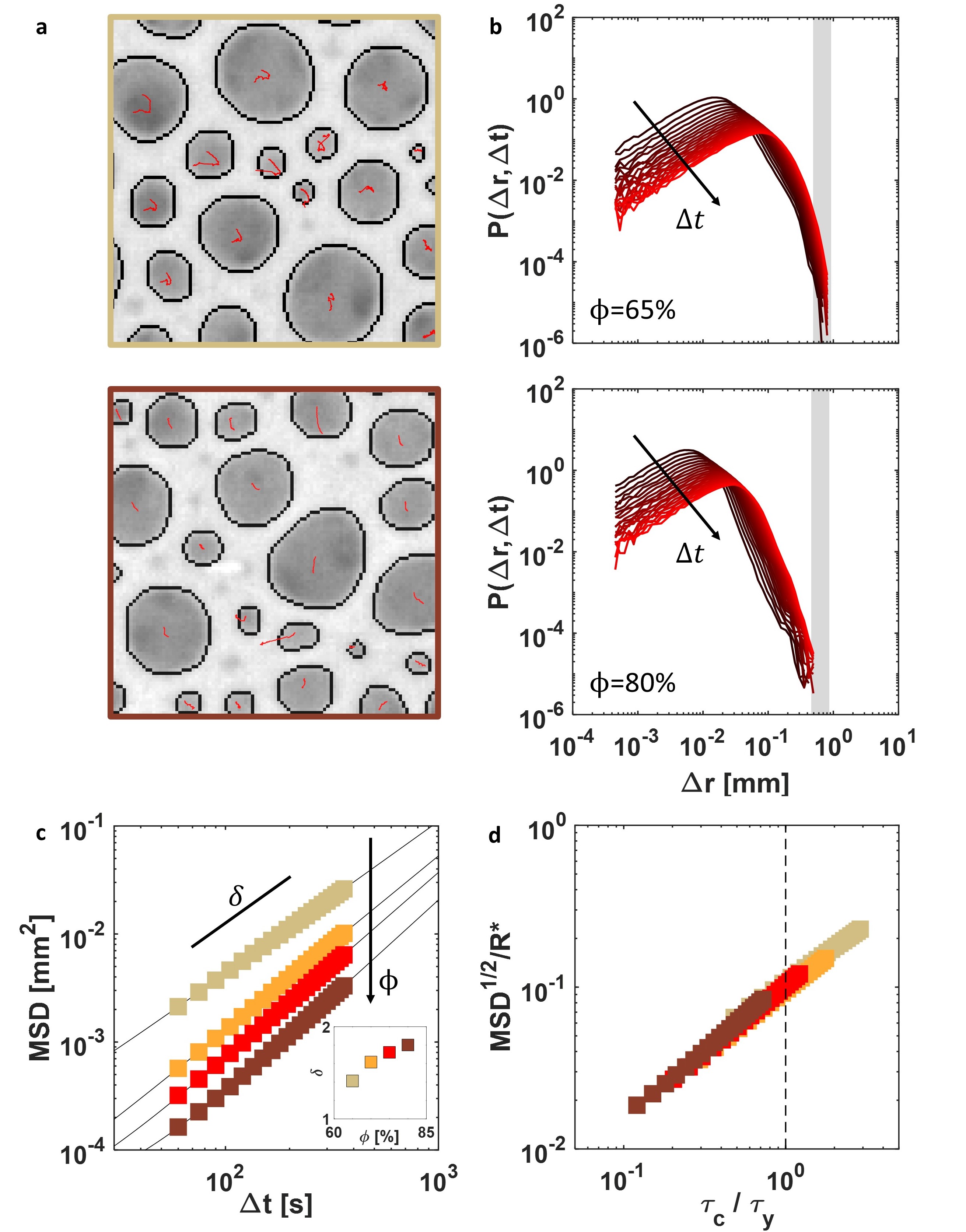}
    \caption{\textbf{Bubble tracking.} (a) Representative examples of bubble trajectories, lasting the whole time window, obtained at $\phi$=65\% (top) and $\phi$=80\% (bottom).
    The frame edge is 4 mm.
    (b) Distribution of bubble displacements at different time delays $\Delta t$ (increasing from 60 s to 360 s with a step of 15 s) for $\phi$=65\% and $\phi$=80\%.
    The vertical gray-shaded bar marks the mean bubble radius $R^*$.
    (c) Bubble mean square displacement as a function of $\Delta t$. MSD $\sim \Delta t^\delta$ for each sample, with $\delta$ increasing with $\phi$, as shown in the inset.
    (d) MSD rescaling. The dashed line marks $\tau_c=\tau_y$.}
\label{fig:FIGURE_3}
\end{figure}

We characterize such differences in the coarsening dynamics by tracking the bubbles.
Since foams are aging, the bubble size, the coarsening rate, and the bubble mobility are all systematically changing over time.
To limit the global bubble growth while tracking the bubble trajectories, we thus restrict the analysis to an image sub-sequence covering a 15-minute time window centered around $t=t^*$, as highlighted in Fig. \ref{fig:FIGURE_1}(e), that ensures a variation in the mean bubble size of less than 15\% for each $\phi$. 
This allows studying bubble dynamics in quasi-stationary conditions \cite{Giavazzi2021}.
Substantially equivalent results obtained for other two non-overlapping time windows, centered around 30 and 45 minutes respectively, are reported in Supplemental Material.

We perform bubble tracking by using TrackMate \cite{TrackMate_1,TrackMate_2} on the segmented foam pictures.
The results obtained at $\phi$=65\% and $\phi$=80\% are reported in Fig. \ref{fig:FIGURE_3}(a), where examples of representative trajectories are shown for both samples.

At $\phi$=65\%, bubbles move persistently in one direction until a sudden deviation occurs, as mirrored by the jumps in the bubble trajectories.
Such abrupt changes of direction are not observed at $\phi$=80\% where, by contrast, trajectories are straighter and shorter (SM, Fig. S7).
In aqueous foams, the loss of directional persistency has been ascribed to the change in the local stress configuration after the occurrence of plastic bubble rearrangements in the surroundings \cite{Giavazzi2021}.
This agrees with our observations: while at low $\phi$ bubbles rearrange during coarsening, at high $\phi$ mutual bubble displacements are almost suppressed (SM, Fig. S9).

To quantify these observations, we compute the probability distribution of bubble displacements $\Delta r$ at different time delays $\Delta t$ (i.e. the self part of the Van Hove correlation function \cite{hansen1986mcdonald}).
For each $\phi$, the distribution exhibits a well-defined peak at each time delay, that systematically shifts towards larger $\Delta r$ as $\Delta t$ is increased, as shown in Fig. \ref{fig:FIGURE_3}(b) for $\phi$=65\% and $\phi$=80\%.
Results for the other values of $\phi$ can be found in Supplemental Material (Fig. S2).
At $\phi$=65\%, the right tail of the distribution decreases as a power law before
dropping at $\Delta r$ around the characteristic bubble size, akin to aqueous foams \cite{Giavazzi2021}.
By contrast, at $\phi$=80\% the distributions decay more steeply and displacements are restricted to smaller length scales.

We compute the bubble mean square displacement MSD$=\langle \Delta r^2\rangle$ at different time delays for each sample.
For this calculation, we discard small bubbles whose trajectories do not cover the whole time window, which correspond to less than 20\% of the total number of detected bubbles.
The MSD dependency on $\Delta t$ is plotted in Fig. \ref{fig:FIGURE_3}(c) for each $\phi$.
The MSD grows asymptotically as a power law MSD $\sim \Delta t^\delta$, with an exponent $\delta$ increasing with $\phi$, heading towards a ballistic-like scaling.

Despite the same coarsening rate, for a fixed $\Delta t$ the MSD shows a tenfold reduction as $\phi$ is increased from 65\% to 80\%, reflecting shorter bubble displacements with increasing emulsion elasticity.
The average bubble size can thus grow in the same way, with very different bubble dynamics.
Changing the continuous phase rheology allows switching from traditional foam coarsening to a new coarsening where mutual bubble displacements are hindered.
The MSD approaching a ballistic scaling at high $\phi$ indeed reflects the bubble tendency to keep moving persistently (SM, Fig. S7), which is a signature of the loss of plastic bubble rearrangements (SM, Fig. S9).

The change in the bubble dynamics can be traced back to the continuous phase rheology.
The emulsion elasticity allows considering the foam continuous phase as a soft elastic solid for small deformations 
\cite{Bey2017}.
The diffusive gas exchange between neighboring bubbles induces a variation in their size, which generates an elastic deformation of the matrix.
Bubble size variations hence give rise to local elastic stresses that tend to restore the initial unstrained structure.
While coarsening, the foam can thus accumulate stress in the matrix due to the change in the bubble packing conditions.
As the strain rate associated to the coarsening process can be written as $\Gamma^* / R^*$ \cite{Sessoms2010, Giavazzi2021}, we estimate the total strain applied to the system in a given time $\Delta t$ as $(\Gamma^* / R^*)\Delta t$, and the total stress accumulated by the foam as $\tau_c = (\Gamma^*/R^*)\Delta t G_0$, where $G_0$ is the elastic modulus of the matrix.
Once this stress overcomes the emulsion yield stress, it can relax via local plastic bubble rearrangements.
Indeed, in order for bubbles to rearrange, the emulsion in the Plateau borders has to yield.
On the other hand, as $\phi$ is increased, the interstitial emulsion can bear higher stresses without yielding.
We thus compare $\tau_c$ with $\tau_y$ by plotting the normalised average displacement MSD$^{1/2}/R^*$ versus the relative stress $\tau_c / \tau_y$.
As shown in Fig. \ref{fig:FIGURE_3}(d), data display a good collapse.
Results obtained at different foam ages also collapse on the same master curve (SM, Fig. S3(d)), meaning that coarsening-driven bubble displacements are set by how much stress the emulsion can bear without yielding.
At high $\phi$ the emulsion between the bubbles is stiff enough to elastically store the stress due to  bubble size variations and counteract bubble rearrangements.
Only when $\tau_c$ equals $\tau_y$ the system has accumulated enough stress to yield the emulsion so that bubbles can rearrange.

Since the mutual bubble displacement is of the order of MSD$^{1/2}$ and the typical inter-bubble distance is of the order of the bubble size $R^*$, we can think of MSD$^{1/2}/R^*$ as a strain.
We hence remark that the threshold $\tau_c=\tau_y$ corresponds to MSD$^{1/2}/R^*\simeq10\%$, which is compatible with the macroscopic strain ($\sim15\%$) needed to yield an equivalent aqueous foam at the same liquid fraction \cite{SaintJalmes1999}.

\begin{figure}[b]
    \centering
    \includegraphics[width=\columnwidth]{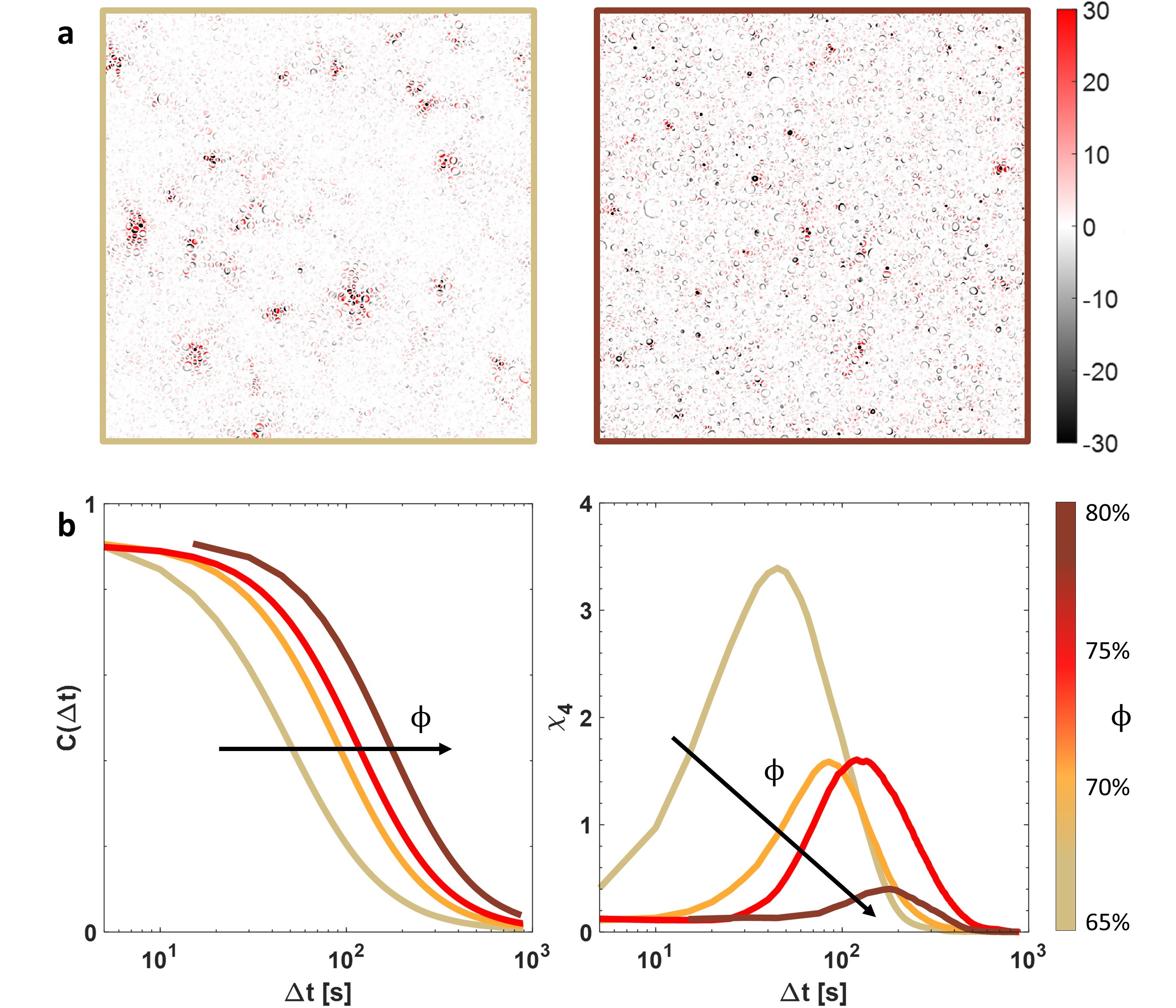}
    \caption{\textbf{Dynamic heterogeneity.} (a) Examples of activity maps at time delays corresponding to the same MSD$\simeq 3 \cdot 10^{-4}$ mm$^2$ for $\phi$=65\% (left) and $\phi$=80\% (right). The edge size is 85 mm. (b) Total mobility $C(\Delta t)$ and dynamic susceptibility $\chi_4$ for each sample.}
\label{fig:FIGURE_4}
\end{figure}

So far, we have only looked at ensemble-averaged dynamic quantities, neglecting any possible spatial fluctuation.
The collapse in Fig. \ref{fig:FIGURE_3}(d) might suggest that at high $\phi$ the dynamics is simply slowed down, so that the system does not accumulate enough stress for rearrangements to occur within the time window considered for quasi-stationarity.
However, if we compare activity maps corresponding to the same MSD (but different time delays), the latter actually show a clear qualitative difference between $\phi$=65\% and $\phi$=80\%, as reported in Fig. \ref{fig:FIGURE_4}(a).
Bubble motion is highly heterogeneous at $\phi$=65\%, as in aqueous foams, with core regions where neighbor switching occurs surrounded by shells of bubbles that adjust at fixed topology.
Such dynamic heterogeneity is lost at $\phi$=80\%, where bubbles globally move more homogeneously.

We characterize spatial dynamic fluctuations by defining a mobility parameter that quantifies how much the bubbles move between times $t$ and $t+\Delta t$.
We define the mobility $c_i(t,\Delta t) = \exp(-|\textbf{r}_i(t+\Delta t) - \textbf{r}_i(t)|^2/d^2)$
for each bubble, with $d$=3\%$R^*$, and compute the total mobility as the ensemble average $C(t,\Delta t) = \langle c_i(t,\Delta t)\rangle$.
From the time fluctuations of the total mobility, we then evaluate the dynamic susceptibility as $\chi_4 = N(\langle C(t,\Delta t)^2\rangle - \langle C(t,\Delta t)\rangle^2)$, which efficiently measures the degree of dynamic heterogeneity \cite{RevModPhys.83.587}.
Results are shown in Fig. \ref{fig:FIGURE_4}(b).
For each sample, $\chi_4$ exhibits a peak at a time delay that grows with $\phi$, mirroring the slowing down of the dynamics.
On the other hand, the decrease of the peak height captures the loss of cooperativity and spatial correlation in bubble rearrangements at high $\phi$.
At high $\phi$, bubbles tend to move persistently, deforming the material without relaxing the accumulated stress via neighbor-switching events.
This eventually leads to the formation of unusual foam structures (SM, Fig. S6), as found with larger bubbles \cite{Guidolin2023}.

In conclusion, we show that continuous phase rheology allows changing the bubble dynamics associated to the pressure-driven coarsening process, showing a novel way for foams to age.
The emulsion yield stress hampers mutual bubble displacements, affecting the coarsening dynamics well before impacting the associated kinetics.
Our results show a decoupling between bubble growth and rearrangements, raising new questions on the role of local bubble dynamics on the global structural evolution.
Moreover, this opens new ways for tuning the internal structure of foams, which are of great interest for all those applications requiring solidification of liquid foams, as bubble organisation directly impacts the mechanical properties \cite{Heitkam2016} and thus the performance of the final material.

The authors acknowledge Véronique Trappe for illuminating discussions.
This work has been partly supported by Associazione Italiana per la Ricerca sul Cancro (AIRC) to C.G. and F.G. (MFAG\#22083).

C.G., E.R., and A.S. conceptualized the experimental study.
F.G. designed the methodology.
C.G. conducted the experiments.
C.G and F.G. analyzed the data. 
C.G., R.C., F.G., and A.S. interpreted the experimental results.
C.G. wrote the manuscript with contributions from all authors.


\section{SUPPLEMENTAL MATERIAL}

\subsection{Coarsening dynamics}

\begin{figure*}[htbp]
    \centering
    \includegraphics[width=\textwidth]{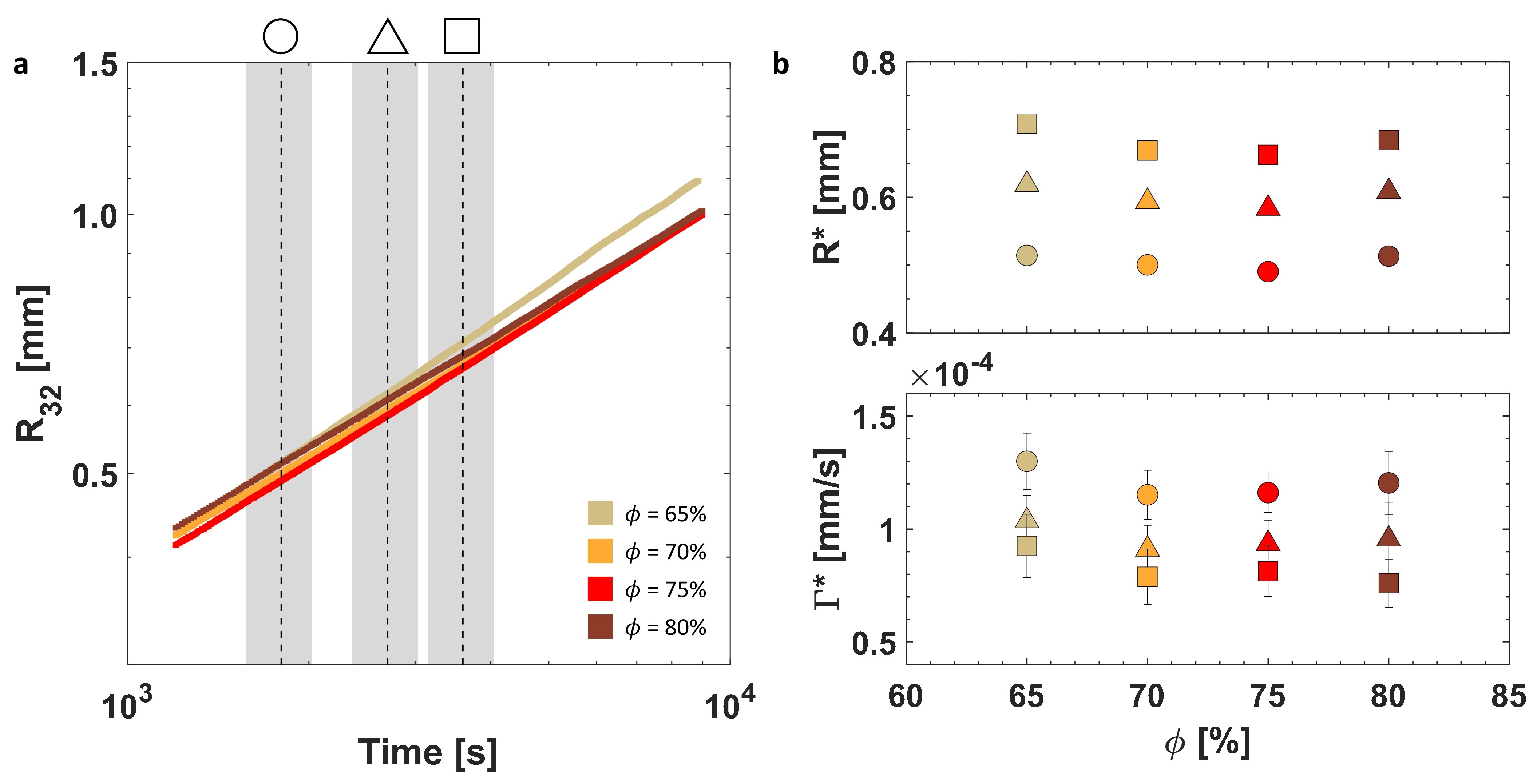}
    \caption{Mean bubble growth. (a) Evolution of the mean bubble radius $R_{32}$ for each sample. The vertical dashed lines mark the times $t^*$ at which we probe the coarsening dynamics. The gray-shaded areas highlight the corresponding non-overlapping time windows.
    (b) Mean bubble radius $R^*$ (top) and coarsening rate $\Gamma^*$ (bottom) obtained for $t^*$ = 30 (circles), 45 (up-pointing triangles), and 60 (squares) minutes. The error bar on the coarsening rate is estimated as the standard deviation of $\Gamma^*$ calculated in 16 different square regions of the sample.}
    \label{fig:SI_R32vsTime_TimeWindows}
\end{figure*}

As foams are evolving over time, we study the bubble dynamics in quasi-stationary conditions by performing bubble tracking on image sub-sequences centered at a certain foam age $t^*$ and covering a $t^*/4$ time window, which ensures a variation of the mean bubble size of less than 15\%.
We consider three different $t^*$ corresponding to 30, 45, and 60 minutes respectively.
Figure \ref{fig:SI_R32vsTime_TimeWindows}(a) shows the time evolution of the Sauter mean bubble radius $R_{32}=\langle R^3\rangle / \langle R^2 \rangle$ for each sample, where the three different $t^*$ are marked as vertical dashed lines.
The gray-shaded areas in the same plot highlight the corresponding time windows considered for the bubble tracking.
The mean bubble size $R^*=R_{32}(t^*)$ and the coarsening rate $\Gamma^* = $ d$R_{32}/$d$t|_{t^*}$ evaluated at each $t^*$ are reported in Fig.\ref{fig:SI_R32vsTime_TimeWindows}(b).

\begin{figure*}[htbp]
    \centering
    \includegraphics[width=\textwidth]{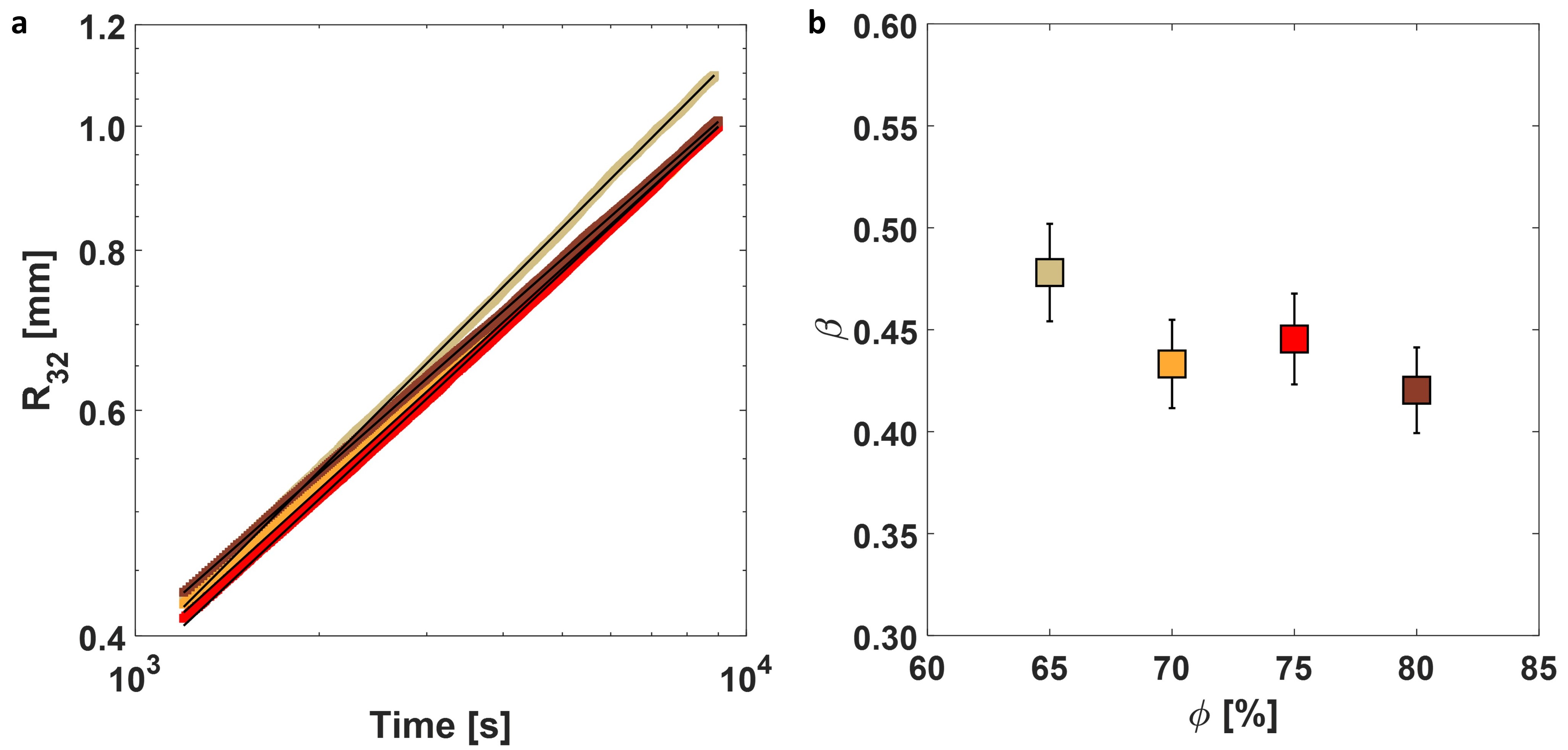}
    \caption{(a) Time evolution of $R_{32}$ fitted with a global power law fit of the kind $R_{32} = \alpha \cdot t^\beta$ (black solid lines). (b) Corresponding exponent $\beta$ as a function of the oil fraction $\phi$.}
    \label{fig:SI_powerlaws_exponents}
\end{figure*}

\begin{figure*}[htbp]
    \centering
    \includegraphics[width=\textwidth]{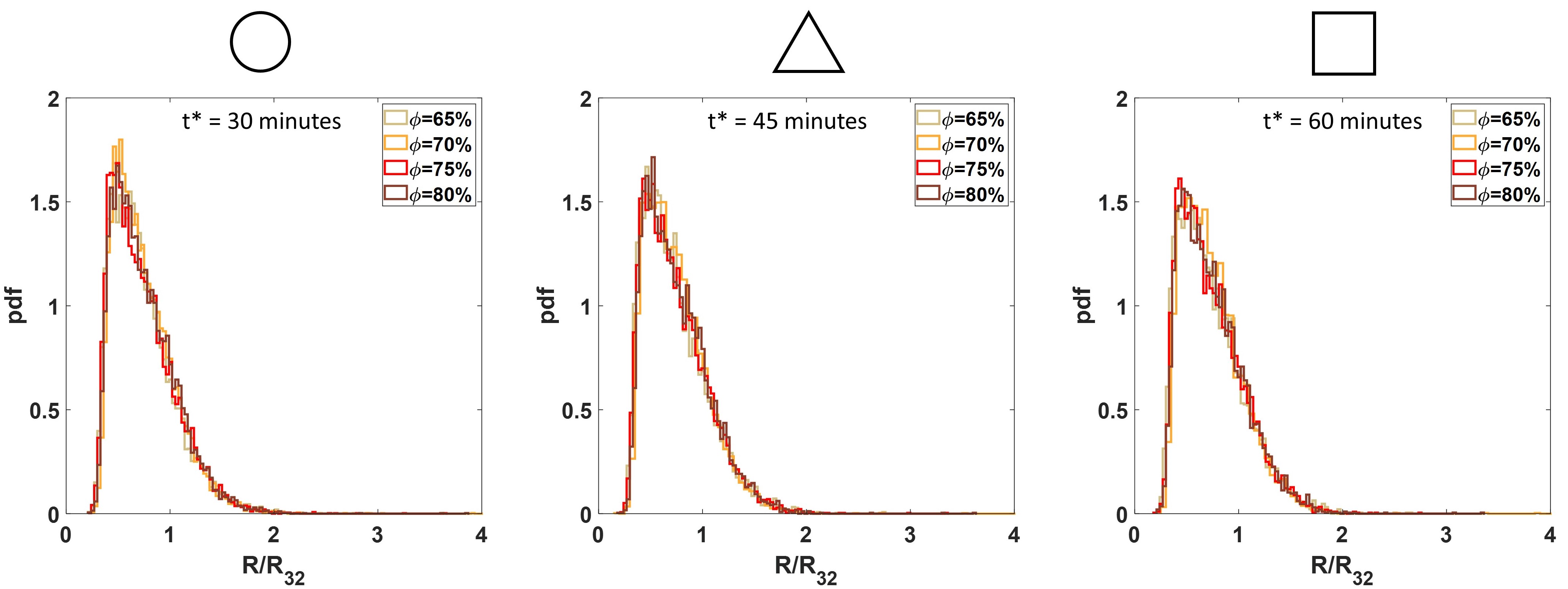}
    \caption{Bubble size distributions at different oil fraction $\phi$ and different foam age $t^*$.}
    \label{fig:SI_SizeDistributions}
\end{figure*}

If we fit the global time evolution of the Sauter mean bubble radius $R_{32}$ with a power law of the kind $\alpha \cdot t^\beta$ for each $\phi$, we obtain the exponents $\beta$ shown in Fig. \ref{fig:SI_powerlaws_exponents}.
However, as our foams are not expected to reach a scaling state, we do not expect a power law growth \textit{a priori}, which is why we decided to reason in terms of the coarsening rate $dR_{32}/dt$ as measured locally in the time windows considered, rather than in terms of power law exponents.
To compare the foam structure between the samples, we calculate the distribution of the bubble radii $R$, normalised by the average value $R_{32}$.
The normalised bubble size distributions are reported in Fig. \ref{fig:SI_SizeDistributions} below for each $\phi$ and each foam age $t^*$ considered.
No significant difference is observed between the size distributions at different $\phi$ at each $t^*$.

\begin{figure*}[htbp]
    \centering
    \includegraphics[width=\textwidth]{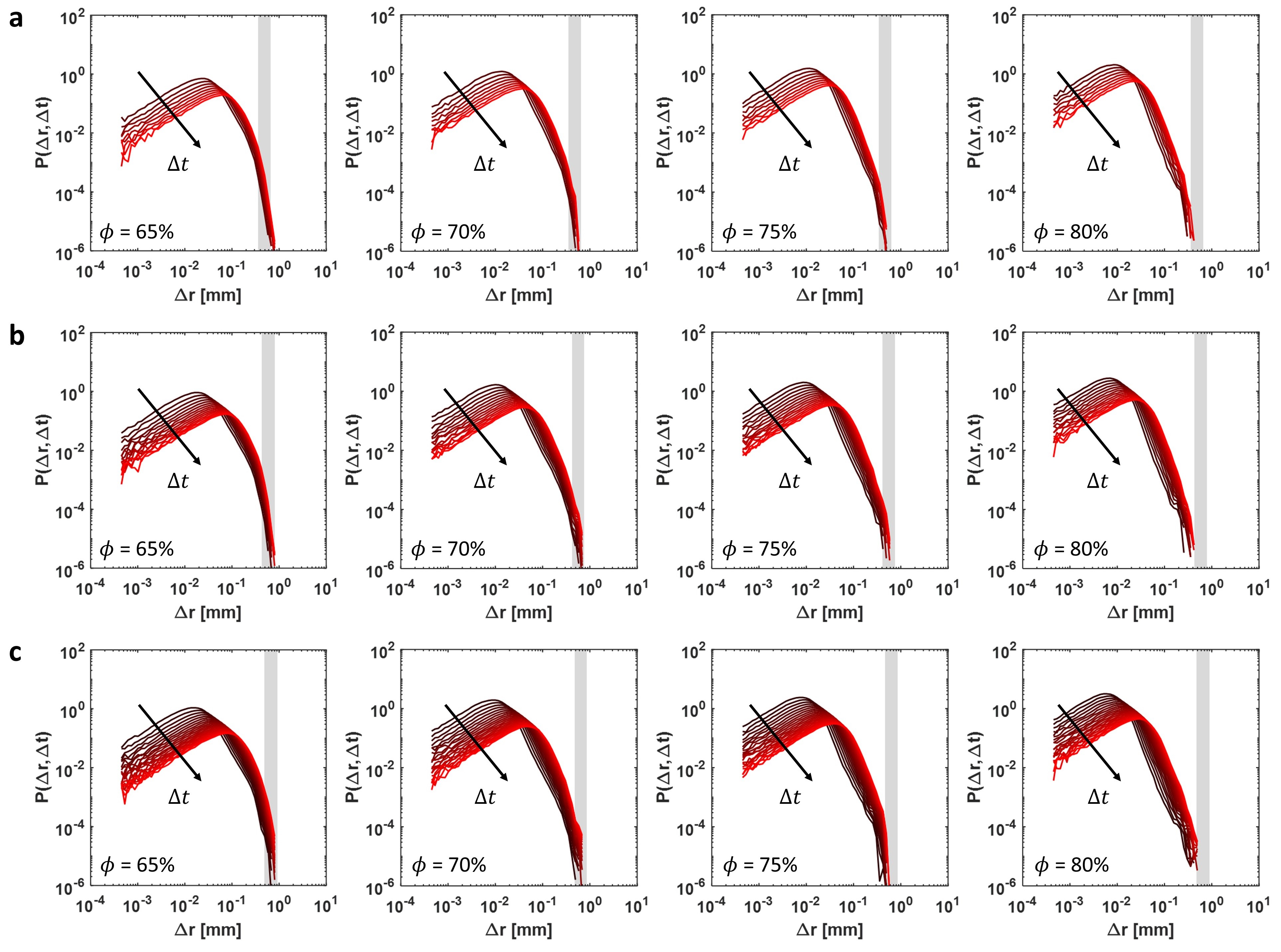}
    \caption{Distributions of bubble displacements for the time window centered at $t^*$ for samples at $\phi$ = 65\%, 70\%, 75\%, and 80\%, as indicated. The results are reported for the time windows centered at $t^* =$ 30 (a), 45 (b), and 60 (c) minutes. The curves correspond to time delays $\Delta t$ increasing from 60 seconds to $t^*/10$ with a step of 15 seconds. The vertical gray-shaded bars highlight the cut-off length corresponding to the mean bubble size $R^*$.}
    \label{fig:SI_Distributions_AllTW}
\end{figure*}

We report in Fig. \ref{fig:SI_Distributions_AllTW} the probability distribution of bubble displacements $P(\Delta r,\Delta t)$ obtained for each sample and each time window.
Let us consider the first time window centered at $t^* =$ 30 minutes.
All distributions exhibit a well-defined peak at each time delay, that systematically shifts to larger $\Delta r$ as $\Delta t$ is increased.
At low $\phi$, the right tail of the distribution decreases as a power law before dropping at $\Delta r$ around the characteristic bubble size, as marked by the vertical gray-shaded bars.
By contrast, as $\phi$ is increased, the decay of the distribution becomes gradually steeper, with bubble displacements stopping at smaller length scales until, at $\phi$=80\%, the distributions no longer reach displacements comparable to the bubble size.
Equivalent results are obtained in the other two time windows and reported in Fig. \ref{fig:SI_Distributions_AllTW}(b,c), showing that the analysis does not depend on the choice of $t^*$ within the experimental time.

\begin{figure*}[htbp]
    \centering
    \includegraphics[width=\textwidth]{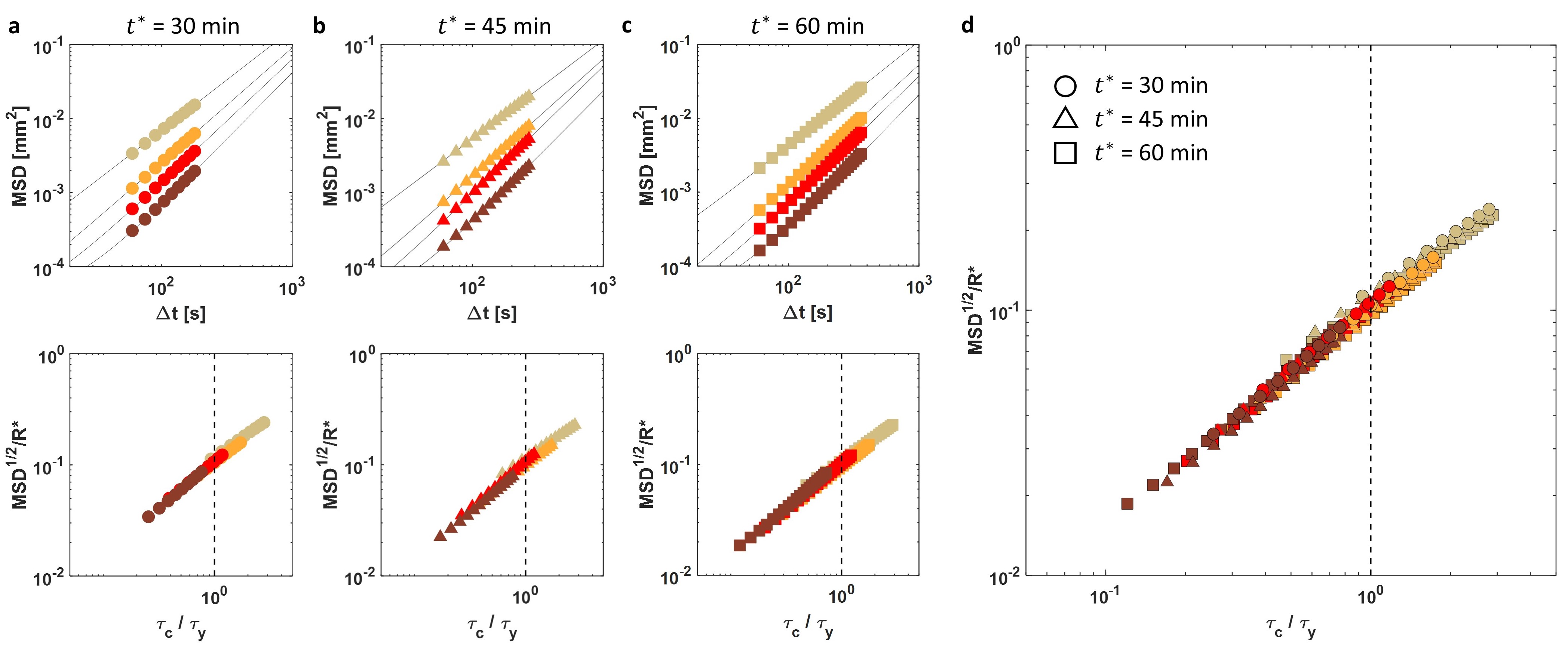}
    \caption{Mean square displacement. (a-c) MSD (top) and its rescaling (bottom) for each sample at different foam ages $t^*$ as indicated in the figure. The black lines are power law fit to the data. The vertical dashed lines mark the threshold $\tau_c = \tau_y$. (d) All rescaled data collapse on the same master curve.}
    \label{fig:SI_MSD_AllTW}
\end{figure*}

From the bubble trajectories, we evaluate the mean square displacement (MSD) at different time delays $\Delta t$.
The $\Delta t$-dependency of the MSD at different $\phi$ is plotted in Fig. \ref{fig:SI_MSD_AllTW} for each time window considered.
For each $\phi$, the MSD grows asymptotically as a power law MSD $\sim \Delta t^\delta$, with exponent $\delta$ increasing with $\phi$ towards a ballistic-like scaling.

As shown in Fig. \ref{fig:SI_MSD_AllTW}, a good collapse of the data is observed for each foam age when plotting the normalised average displacement MSD$^{1/2}/R^*$ versus the relative stress $\tau_c / \tau_y$, where $\tau_c = (\Gamma^*/R^*)\Delta t G_0$ is the total stress accumulated in the system due to coarsening, and $\tau_y$ is the emulsion yield stress.
We remark that the results obtained at different foam ages collapse on the same master curve as shown in Fig. \ref{fig:SI_MSD_AllTW}(d).

\subsection{Gravitational drainage.}

Three dimensional foams are subject to gravitational drainage: the liquid between the bubbles gradually flows downwards because of gravity, making the foam drier on top and wetter at the bottom.
Drainage makes probing 3D foam coarsening very challenging, which is why studies on moderately wet foams typically resort to quasi-2D configurations \cite{glazier1987dynamics,stavans1989soap,chieco2021experimentally} or microgravity conditions \cite{pasquet2023aqueous,pasquet2023coarsening,galvani2023hierarchical}.
In our case, since we monitor the foam evolution from the top, drainage could result in an overestimate of the coarsening rate.
However, the emulsion yield stress allows delaying the foam gravitational drainage, providing a time range in which the liquid fraction can be considered homogeneous inside the sample and thus the coarsening of the top surface bubble layer is representative of the bulk evolution.

Gravitational drainage is expected to become significant above a critical bubble size at which the buoyancy force per unit area exerted on the bubble overcomes the emulsion yield stress, namely when $\rho g R / 3 > \tau_y$ \cite{Goyon2010_PRL_drainage}.
Since the emulsion yield stress depends on its oil fraction, coarsening can be safely studied without substantial gravitational effects up to a critical bubble size, estimated as $R_d = 3\tau_y / \rho g$, which increases with $\phi$.

Our setup does not allow probing coarsening in the absence of gravitational drainage with emulsion oil fractions below jamming or with an aqueous continuous phase at similar bubble sizes and liquid fractions.
To avoid mixing coarsening with gravitational effects, we consider a range of $\phi$ between 65\% and 80\%, and we stop the image acquisitions after a couple of hours.
Indeed, we estimate the critical bubble radius for the onset of drainage to be of the order of hundreds of microns for the lowest oil fraction investigated $\phi=65\%$ and to be several millimeters at $\phi>65\%$.
Therefore, while the slight change of curvature at the end of the $R_{32}(t)$ curve observed at $\phi=65\%$ can be reasonably ascribed to the onset of drainage, the measurements on all samples at $\phi>65\%$ stop at bubble radii well below the critical size.
However, estimating the bubble size from the foam skeleton allows reducing possible artifacts due to changes in the liquid fraction.
In addition, the appearance of the bottom bubble layer of the foam at the end of the image acquisition has been checked for each $\phi$.
Some gravity-induced vertical size segregation inside the samples can occur even at higher $\phi$ because of bubble polydispersity, with some large bubbles rising to the top of the sample, but their contribution to the overall kinetics and dynamics is negligible.

\subsection{Late-stage foam structure}

Our experiments show that continuous phase stiffness impacts the coarsening dynamics well before the appearance of clear changes in the foam structure.
However, bubble inability to plastically rearrange can be at the origin of the onset of heterogeneous coarsening observed at millimetric bubble sizes in quasi-2D foams \cite{Guidolin2023}.
Lack of emulsion redistribution inside the foam can indeed play a crucial role in the initial stage, before inhomogeneities in the foam structure, and thus in its local mechanical properties, arise and further exacerbate spatially heterogeneous bubble growth.

\begin{figure*}[htbp]
    \centering
    \includegraphics[width=\textwidth]{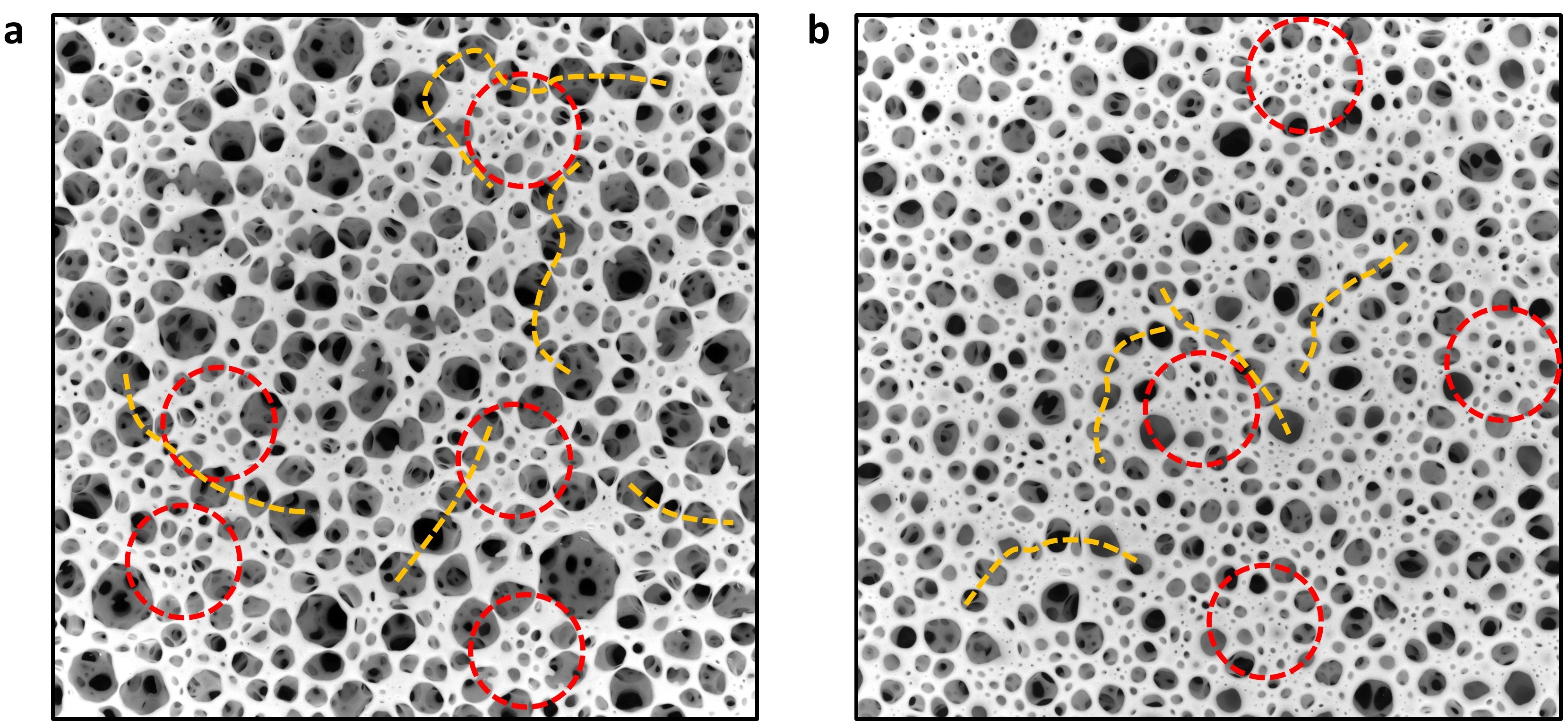}
    \caption{Late-stage foam structure at $\phi =$ 80\%. The photos show the appearance of the top (a) and the bottom (b) of the sample after 23 hours. Despite the occurrence of coalescence events and the onset of gravitational drainage, the pictures clearly show an atypical bubble pattern induced by heterogeneous coarsening. Examples of clusters of small bubbles are highlighted with red circles, while dashed yellow lines indicate examples of chains of large bubbles. The edge size is 11 cm.}
    \label{fig:SI_FoamStructure_phi80}
\end{figure*}

To probe this effect in 3D foams, we let the sample at $\phi=80\%$ age to qualitatively monitor from the top its structure evolution.
At the end of the image acquisition, the foam cell has been turned upside down to take a picture of the bottom, to visually check on the effect of drainage.
The late-stage foam appearance is shown in Fig. \ref{fig:SI_FoamStructure_phi80}.
One can recognise the effect of gravitational drainage, visible from the difference in the liquid fraction between the top and the bottom of the sample, as well as the traces of the occurrence of film-rupture events on the top bubble layer.
The combination of three different mechanisms of foam destabilisation (coarsening, coalescence, and drainage) prevents any quantitative assessment on bubble growth and dynamics at this stage.
However, we can see how the foam exhibits an atypical coarsening-induced structure, similar to the ones observed in quasi-2D configurations \cite{Guidolin2023}, with islands of small bubbles enclosed by chains of larger bubbles, as highlighted with guides to the eye.

\subsection{Persistence length}

\begin{figure*}[htbp]
    \centering
    \includegraphics[width=\textwidth]{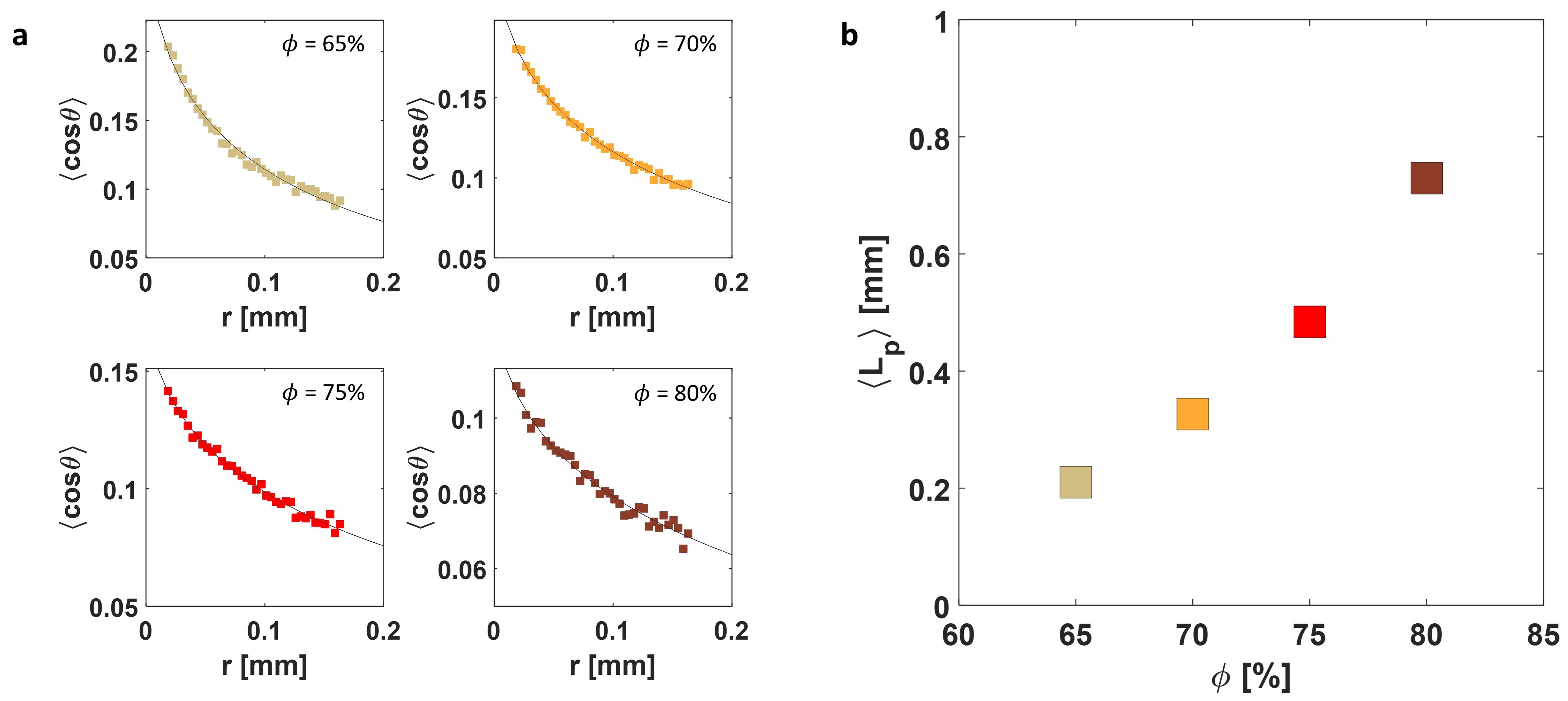}
    \caption{Persistence length for $t^*=60$ minutes. (a) Correlation curves obtained for each $\phi$. The solid black line represents the stretched exponential fit to the data. (b) Persistence length $\langle L_p \rangle$ obtained from the fit.}
    \label{fig:SI_PERSISTENCE_LENGTH}
\end{figure*}

From the bubble trajectories, we quantify the associated persistence length as follows.
In analogy with polymers \cite{doi1988theory}, we consider the orientational correlation function $\langle \cos\theta(r)\rangle$, where $\theta(r)$ is the angle between the vectors tangent to the trajectory in two points separated by a portion of trajectory of length $r$. For a semi-flexible polymer, the orientational correlation function is expected to fall off exponentially $\langle \cos\theta(r) \rangle = e^{-r/L_p} $, where $L_p$ is the persistence length and the angled brackets denote the average over all starting positions along the trajectory.\\
For this calculation, we consider only trajectories covering the whole time window and identify the direction of tangent vectors with the one of the segment connecting two consecutive points of the trajectory.
The correlation curves, averaged over all the bubbles, are shown in Fig. \ref{fig:SI_PERSISTENCE_LENGTH}(a).
By fitting these decays with a stretched exponential of the kind $\langle cos\theta(r) \rangle = \alpha \cdot \exp(-(r/L_p)^\beta)$ we can extract the persistence length $L_p$, which is then corrected for the stretching exponent to retrieve the mean value as $\langle L_p \rangle = \frac{L_p}{\beta} \Gamma(\frac{1}{\beta})$, where $\Gamma$ is the gamma function.
The amplitude term $\alpha<1$ is introduced to account for the drop of orientational correlation between two consecutive displacements due to the localization error, while the same value $\beta=0.5$ for the stretching exponent is used for all samples. 
The persistence length $\langle L_p \rangle$ for $t^*=60$ minutes is reported in Fig. \ref{fig:SI_PERSISTENCE_LENGTH}(b) as a function of $\phi$. The corresponding best fitting values for $\alpha$ are $(0.30 , 0.25, 0.18, 0.13) \pm 0.08$ for $\phi=(65\%, 70\%,75\%,80\%)$, respectively.
We can see that $\langle L_p \rangle$ grows with $\phi$, confirming that bubble trajectories become straighter with increasing oil fraction, namely bubbles tend to move persistently.

\subsection{Bubble rearrangements}

\begin{figure*}[htbp]
    \centering
    \includegraphics[width=\textwidth]{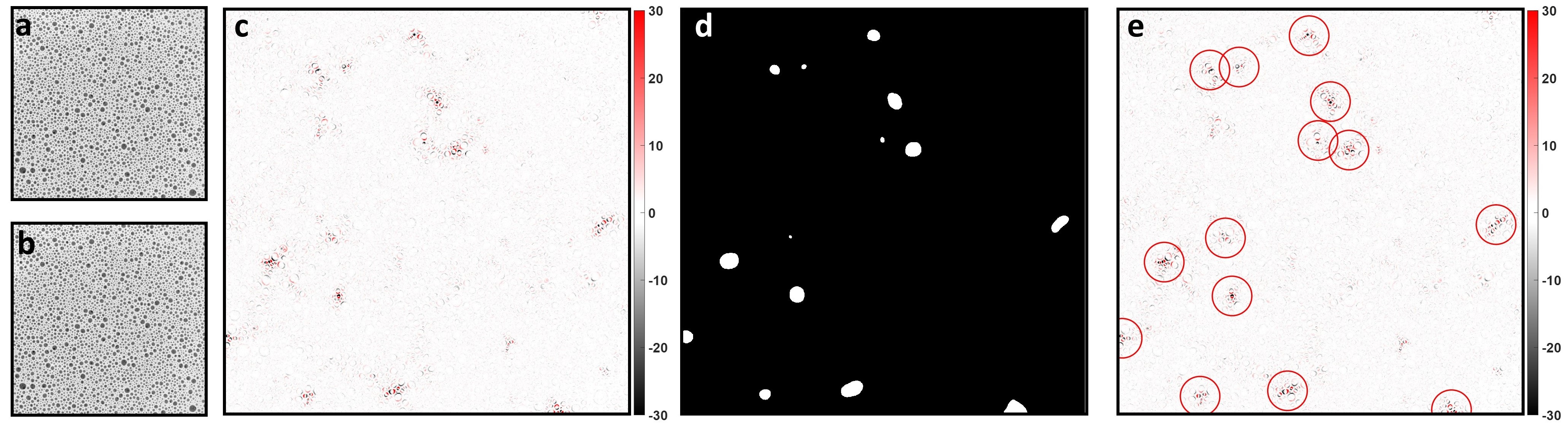}
    \caption{Detection of bubble rearrangements. (a,b) Source and target foam images separated by a time delay $\Delta t$ = 15 seconds. (c) Corresponding activity map, that is the image difference displayed with a custom colormap. (d) Output image after thresholding, showing the blobs corresponding to the spots of highest activity in the foam, as shown in (e) where the detected active regions are highlighted with red circles.}
    \label{fig:SI_BLOB_DETECTION}
\end{figure*}

\begin{figure*}[htbp]
    \centering
    \includegraphics[width=\textwidth]{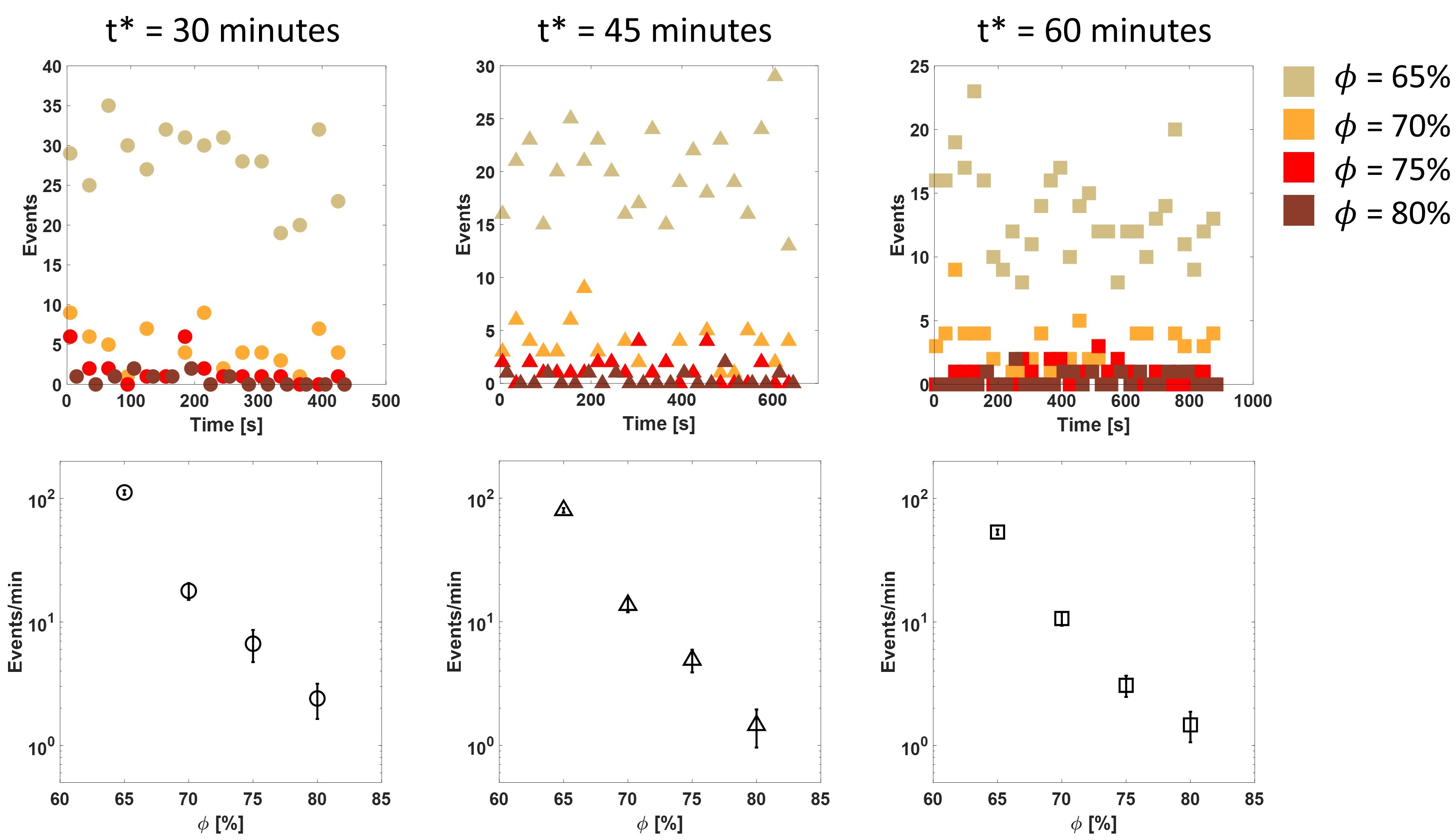}
    \caption{Bubble rearrangements. Top row: raw number of events detected for each $\phi$ at each foam age $t^*$. Bottom row: corresponding rate of events as a function of $\phi$.}
    \label{fig:SI_BLOB_counting}
\end{figure*}

An exact quantification of the rate of bubble rearrangements from local topological changes in the foam skeleton is extremely sensitive to errors in bubble detection, which likely occurs in our three-dimensional foam images.
However, we can get a fully automatic and robust estimate of the number of bubble rearrangements from the activity maps as follows.\\
We identify the bubble rearrangements as the local regions of highest activity in the foam.
We compute the difference between images separated by a time delay of 15 seconds, as shown in Fig. \ref{fig:SI_BLOB_DETECTION}(a-c).
Since the coarsening rate is the same between the samples, we fix the same time interval for each $\phi$.
This is the same procedure used to get the activity maps shown in Fig. 2 in the main text.
We then calculate the square of the image difference and apply a Gaussian filter of standard deviation $\simeq$ 0.8 mm before thresholding.
An example of the output image is shown in Fig. \ref{fig:SI_BLOB_DETECTION}(d), where we can recognize white blobs on a dark background, corresponding to the regions of highest activity.
The time separation between images is chosen to prevent the merging of blobs corresponding to different events, ensuring a correct counting of the number of events.
We repeat this procedure for all pairs of source and target images separated by the same time delay $\Delta t=$ 15 s within the time windows centered at foam ages $t^*$ = 30, 45, and 60 minutes.
To prevent double counting of the same event on consecutive image pairs, we consider only source images separated by twice the time delay $\Delta t$.
The raw number of events counted for each sample at each foam age is shown in Fig. \ref{fig:SI_BLOB_counting}(top row).
We can see a marked reduction in the number of detected events with increasing $\phi$, going from tens to only a few events as we switch from $\phi$  = 65\% to $\phi$ = 80\%.
At the lowest $\phi$, we can also recognize a slight decrease in the number of events over time, consistently with the slowing down of the coarsening kinetics.
As the mean bubble size $R^*$ and the image size do not change with $\phi$, the number of bubbles is also the same so that we can directly compare the rate of events.
We then calculate the average rate of rearrangements by dividing the raw number of events by the lag time and averaging over the whole time window.
The results are reported in Fig. \ref{fig:SI_BLOB_counting}(bottom row), where we can see that, despite the same coarsening rate, the rate of bubble rearrangements is strongly reduced with increasing oil fraction $\phi$ at each foam age $t^*$.\\

\subsection{Coarsening movies}
We provide four movies showing the coarsening dynamics at different $\phi$, corresponding to the 15-minute time window centered around $t^* = 60$ minutes.
Consecutive frames are separated by a time delay of 15 seconds and displayed at a playback speed of 5 fps.
The size of each frame edge is 85 mm. 
Movies SM1, SM2, SM3, and SM4 correspond to $\phi$ = 65\%, 70\%, 75\%, and 80\% respectively.


\end{document}